\documentclass[aps, twocolumn,floatfix,pre,showpacs, superscriptaddress]{revtex4-2}

\usepackage{amssymb,amsmath,amsfonts}
\usepackage{graphicx,graphics}
\usepackage{epsfig}
\usepackage{epstopdf}
\usepackage{amsfonts}
\usepackage{bm,bbm}
\usepackage{amssymb,amsmath,amsfonts}
\usepackage{mathrsfs}
\usepackage{calc}
\usepackage{epsfig}
\usepackage{float}
\usepackage{xcolor}
\usepackage{epstopdf}
\usepackage{bm,amssymb,amsmath}
\usepackage{graphicx,color}

\usepackage{ulem}
\usepackage{xspace}

\usepackage[colorlinks=true, citecolor=teal, urlcolor=purple,linkcolor=purple]{hyperref}

\begin{document}

\title{Time-energy tradeoff in stochastic resetting using optimal control}

\author{R\'emi Goerlich}
\email{remigoerlich@tauex.tau.ac.il}
\affiliation{Raymond \& Beverly Sackler School of Chemistry, Tel Aviv University, Tel Aviv 6997801, Israel}
\affiliation{Institut für Theoretische Physik II - Weiche Materie, Heinrich-Heine-Universität Düsseldorf, D-40225 Düsseldorf, Germany}

\author{Kristian St\o{}levik Olsen}
\affiliation{Institut für Theoretische Physik II - Weiche Materie, Heinrich-Heine-Universität Düsseldorf, D-40225 Düsseldorf, Germany}

\author{Hartmut L\"{o}wen}
\affiliation{Institut für Theoretische Physik II - Weiche Materie, Heinrich-Heine-Universität Düsseldorf, D-40225 Düsseldorf, Germany}

\author{Yael Roichman}
\email{roichman@tauex.tau.ac.il}
\affiliation{Raymond \& Beverly Sackler School of Chemistry, Tel Aviv University, Tel Aviv 6997801, Israel}
\affiliation{Raymond \& Beverly Sackler School of Physics and Astronomy, Tel Aviv University, Tel Aviv 6997801, Israel}

\date{\today}

\begin{abstract}
Stochastic resetting is a driving mechanism that is known to minimize the first passage time to reach a target, at the cost of energy expenditure.
The choice of the physical implementation of each resetting event determines the tradeoff between the acceleration of the search process and its energetic cost.
Here, we present an optimal transport protocol that balances the duration and the energetic cost of each resetting event.
This protocol drives a harmonically trapped Brownian particle between two equilibrium states within a finite time and with minimal energetic cost.
An explicit comparison with other types of finite-time protocols further shows its specific thermodynamic properties.
Its cost is both a lower bound on the cost of unoptimized shortcut protocols and an upper bound on the cost of optimal protocols which do not ensure final equilibrium.
When applying the optimal transport protocol to implement stochastic resetting, a single lower time-energy bound is reached: this protocol allows to reach the best tradeoff between energetic cost and search time.
\end{abstract}

\maketitle

\textit{Introduction} ---
Stochastic resetting (SR), initially introduced in the context of diffusive motion, involves interrupting a stochastic process at random intervals and resetting it to the origin \cite{Evans2011, Evans2020, kundu2022stochastic, Gupta2022}. Remarkably, SR was shown to substantially reduce the mean first-passage time (MFPT) required for a diffusing particle to reach a specified target \cite{Evans2011, evans2013optimal}. However, this benefit is accompanied by an energetic cost, which has been the subject of intensive research in recent years \cite{Fuchs2016, Busiello2020, Gupta2020_thermo, TalFriedman2020, Mori2022, goerlich2023experimental, olsen2024thermodynamic, gupta2025thermodynamic}.
Recent experimental implementations of SR have utilized confining potentials to guide diffusing particles back to their origins \cite{TalFriedman2020, Besga2020, Besga2021, goerlich2023experimental}.
Naturally, such experiments rely on finite-time manipulations that increase the MFPT to the target \cite{evans2018effects, pal2019invariants, gupta2020stochastic, bodrova2020resetting, gupta2021resetting, gupta2022work}.
To reduce the energetic cost of resetting, slow protocols were used, which further increase the MFPT \cite{goerlich2023experimental, olsen2024thermodynamic, gupta2025thermodynamic}.
This underscores the importance of comprehensive control over both the duration of resetting events and their associated thermodynamic costs.

Here, we propose to meet this challenge with tools of a distinct field of research: control theory in finite-time thermodynamics \cite{bechhoefer2021control, alvarado2025optimal, schmiedl2007efficiency, esposito2010efficiency, aurell2012refined}.
This allows us to tailor the return protocol driving the particle back to the origin in SR,  minimizing energetic cost as well as the duration of the search process.
In the field of finite-time thermodynamics, the control of small dissipative systems has been addressed in the past using different approaches, which possess distinct features.
The first point of view focuses on accelerating the relaxation of a system to equilibrium. Originating from the study of shortcuts to adiabaticity in quantum systems \cite{guery-odelin_shortcuts_2019}, these protocols were later applied under the name of swift state-to-state transformation (\textbf{SST}) to mesoscopic systems subjected to thermal noise \cite{martinez_engineered_2016, guery-odelin_shortcuts_2019, guery-odelin_driving_2023, baldovin2023control, Raynal2023}. For these \textit{constrained yet unoptimized} protocols, the final state is prescribed, but no thermodynamic optimization is enforced, often resulting in significantly more dissipation.
In contrast, the field of optimal control seeks to minimize dissipation without any constraint on the final state of the system \cite{Schmiedl2007, loos2024universal, aurell2011optimal, proesmans2020finite, proesmans2020optimal}. These \textit{unconstrained optimal control} (\textbf{OC}) protocols let the system evolve naturally according to its intrinsic relaxation dynamics once the manipulation protocol has ended, only reaching equilibrium asymptotically.

\begin{figure}
	\centerline{\includegraphics[width=0.95\linewidth]{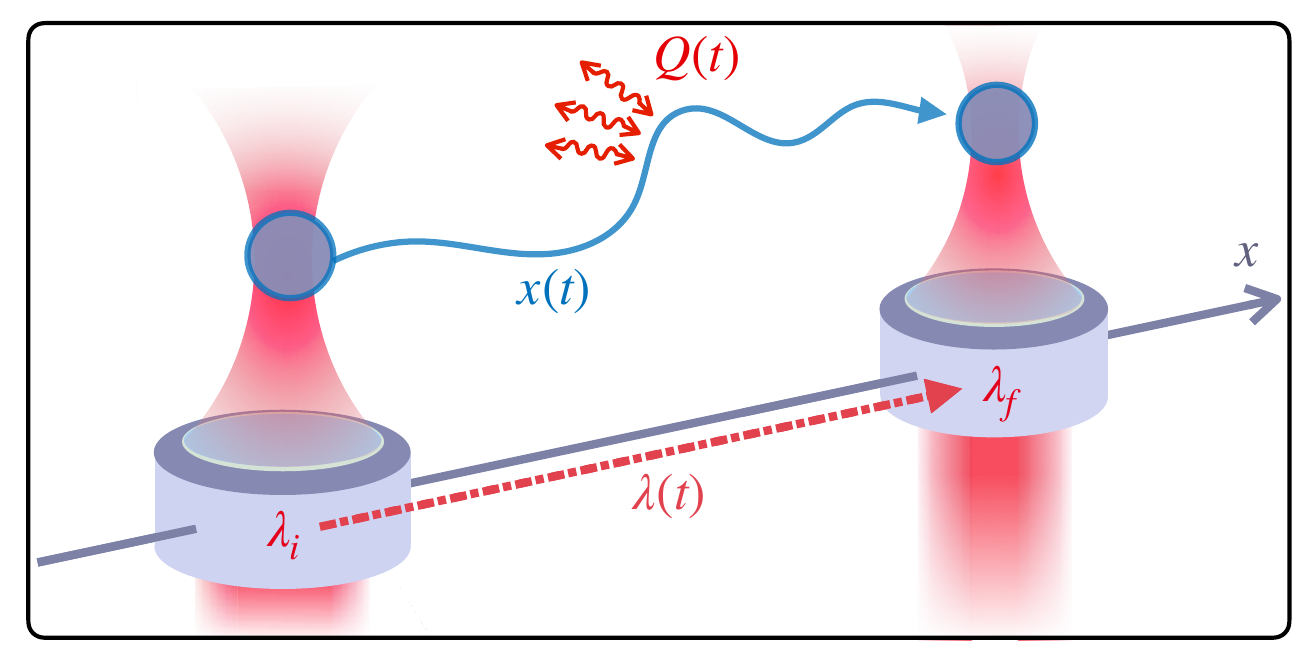}}
	\caption{Optimal protocol $\lambda(t)$ transporting a harmonically trapped Brownian particle from an equilibrium state centered in $\lambda_i$ to a new equilibrium state, centered in $\lambda_f$, within a finite time $t_f$ while minimizing the energetic cost of the transformation. This is achieved by a time-energy optimization procedure \cite{Rosales2020, Pires2023, oikawa2025experimentally}, which combines both a swift equilibration \cite{martinez_engineered_2016} with thermodynamically optimal solutions \cite{Schmiedl2007}.}
	\label{fig:Schema}
\end{figure}

Naturally, these two research directions led to the development of a third class of control protocols, \textit{constrained optimal transport} (\textbf{OT}) protocols as illustrated in Fig.~\ref{fig:Schema}.
These protocols are derived by minimizing dissipation under the constraint of reaching thermal equilibrium at the end of the protocol \cite{aurell2011optimal}.
Such protocols were derived and implemented experimentally for transitions in trap stiffness both in overdamped \cite{Rosales2020} and inertial regimes \cite{baldovin2024optimal}, as well as for transitions involving temperature changes \cite{Pires2023}.

The optimal time-energy tradeoff achieved by the constrained OT protocol is pivotal in any operation where both duration and thermodynamic cost are crucial, for example, cyclic Brownian motors \cite{blickle2012realization, martinez2016brownian, schmiedl2007efficiency} or SR as we demonstrate here.\\

\textit{Outline} ---
Using a Lagrangian-based constrained minimization, we provide a direct derivation of an OT protocol which transports a Brownian particle using a moving harmonic potential (Fig.~\ref{fig:Schema}).
The protocol obtained has a very simple profile, which replicates results from optimal transport geometry \cite{ito2018stochastic, nakazato2021geometrical, zhong2024beyond}.
By comparing this protocol to OC protocols \cite{Schmiedl2007} and unoptimized SST methods \cite{martinez_engineered_2016}, we show that its energetic cost is the upper bound for the former and the lower bound for the latter. This double limiting role reflects the optimal tradeoff between duration and dissipation.
Finally, we apply the OT protocol to search under stochastic resetting, where there is a tradeoff between the energetic cost of resetting and the time to reach the target.
The use of the OT protocol in this inherent tradeoff establishes a performance bound for alternative finite-time resetting implementations and constitutes the main finding of this study.\\

\textit{Derivation of the optimal protocol \,---\,}
We consider a Brownian particle trapped in a harmonic potential $V(x,\lambda) = \frac{1}{2} \kappa (x - \lambda)^2$, centered in $\lambda$ with stiffness $\kappa$.
The dynamics of the particle follows an overdamped Langevin equation
\begin{equation}
    \dot x(t) = - \omega_0 \left(x(t) - \lambda(t)\right) + \sqrt{2 D}\xi(t)
    \label{eq:Langevin}
\end{equation}
where $\omega_0 = \kappa / \gamma$, with $\gamma$ denoting the Stokes friction coefficient. The diffusion coefficient is given by $D = k_B T / \gamma$, where $k_B$ is Boltzmann’s constant and $T$ is the absolute temperature.
We define $u(t) \equiv \langle x(t) \rangle$ as the mean position, averaged over the noise $\xi(t)$, which obeys the following simple differential equation
\begin{equation}
    \dot u(t) = - \omega_0 \left(u(t) - \lambda(t)\right).
    \label{eq:EOM}
\end{equation}
Using this simple equation of motion, we derive the optimal protocol $\lambda_{\rm OT}(t)$ that drives the mean position from the initial value $u_i = \lambda_i$ at time $t_i = 0$ to a final position $u_f = \lambda_f$ within a finite time $t_f$, while minimizing the associated energetic cost.

The derivation, detailed in Appendix~\ref{App:Optimal}, is based on the functional $\mathcal{J} = \int_0^{t_f} \left( \dot W + \mu [\dot u - \omega_0(\lambda-u)]\right) dt$.
The first term in the functional corresponds to the mean work cost of the transformation, evaluated with the tools of stochastic thermodynamics $W = \int\dot W dt = \int \kappa (\lambda - u) \dot \lambda dt$ \cite{Sekimoto1998, SekimotoBook, seifert2012}.
The second term imposes the equation of motion Eq.~(\ref{eq:EOM}) with a Lagrange multiplier $\mu$.
Minimizing $\mathcal{J}$ using the Euler-Lagrange equations yields $\ddot u = 0$.
Imposing the boundary conditions for $u$ directly gives the optimal transport path $ u_{\rm OT}(t) = \lambda_i + \Delta \lambda t / t_f $ where $\Delta \lambda = \lambda_f - \lambda_i$.
Finally, inserting it into the equation of motion gives the optimal transport protocol
\begin{equation}
    \lambda_{\rm OT}(t) = \lambda_i + \Delta \lambda\frac{ \omega_0 t + 1}{\omega_0 t_f}.
    \label{Eq:OptProtocol}
\end{equation}
This protocol, obtained here by a direct Euler-Lagrange minimization, also corresponds to the optimal transport map for this Gaussian system in the space spanned by the $L^2$-Wasserstein metric \cite{nakazato2021geometrical}.
It naturally takes the form $\lambda_{\rm OT}(t) = u_{\rm OT}(t) + l$: the sum of the geodesic path $u_{\rm OT}(t)$ and a counterdiabatic term $l = \Delta\lambda / \omega_0 t_f$ that induces discontinuous jumps \cite{zhong2024beyond} (see Appendix \ref{App:Geom} for details).\\

\begin{figure}[htb]
	\centerline{\includegraphics[width=1\linewidth]{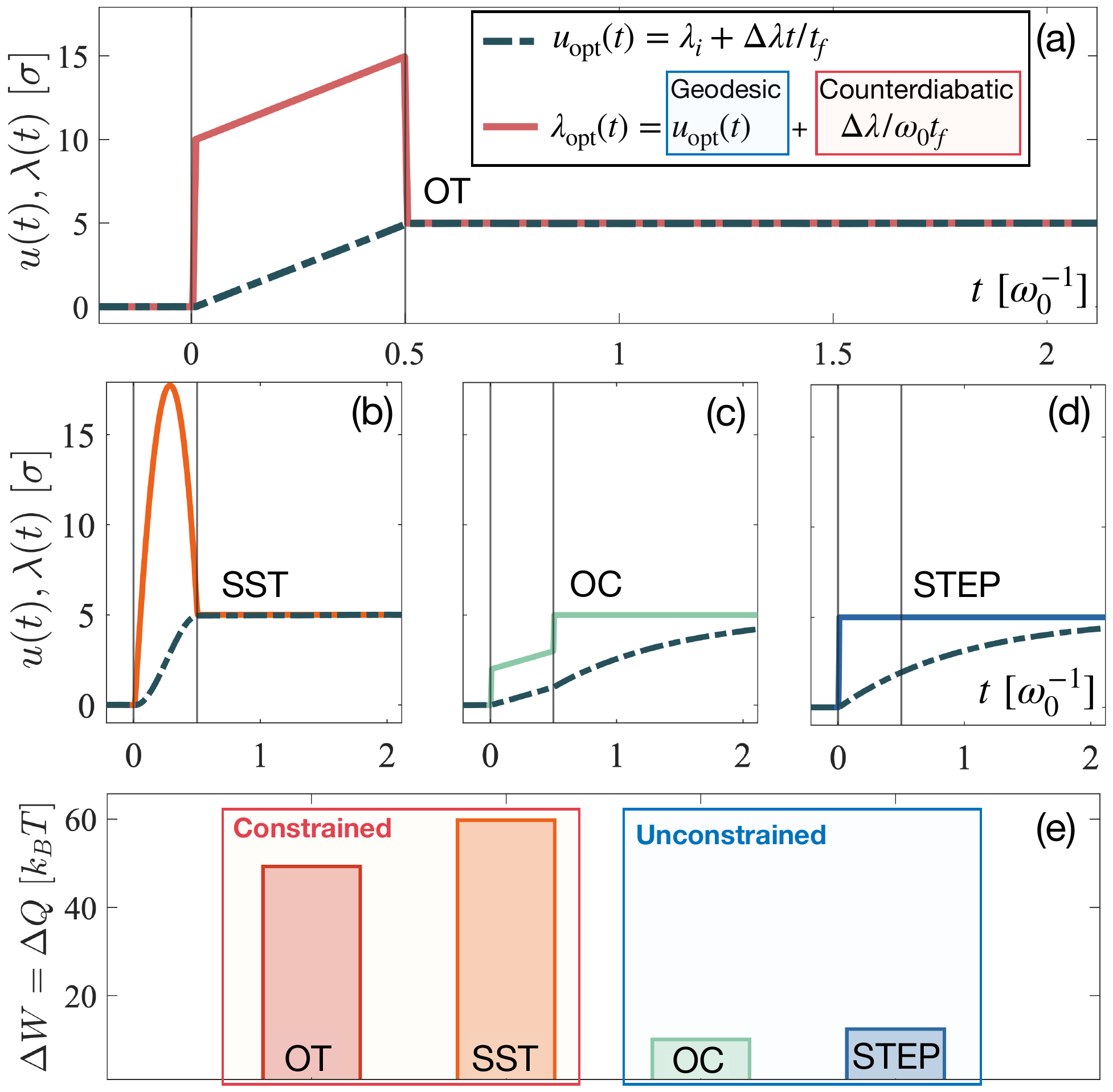}}
	\caption{(a) Optimal Transport (\textbf{OT}) protocol $\lambda_{\rm OT}(t)$ (red line) in units of the standard deviation $\sigma = \sqrt{k_B T/\kappa}$ of the trapped particle, as a function of time, in units of the natural relaxation time $\omega_0^{-1}$.
    The mean position $u(t)$ (gray-blue dash-dotted line) reaches equilibrium $u_{\rm OT}(t_f) = \lambda_f$ exactly in the prescribed time $t_f = \omega_0^{-1} / 2$ (vertical red line).
    In the legend, we emphasize the geodesic and counterdiabatic components of the protocol \cite{zhong2024beyond}.
    (b) Constraint but unoptimized Swift State-to-state Transformation (\textbf{SST}) protocol $\lambda_{\rm SST}(t)$ (orange solid line) and associated $u(t)$.
    (c) Unconstrained Optimal Control (\textbf{OC}) protocol $\lambda_{\rm OC}(t)$ (light green line) and associated $u(t)$.
    (d) Unconstrained and unoptimized abrupt (\textbf{STEP}) protocol $\lambda_{\rm STEP}(t)$ (blue line) and associated $u(t)$.
    (e) Total work cost $\Delta W$ corresponding to each protocol.}
	\label{fig:ProtocolsAndEnergetics}
\end{figure}

\textit{Thermodynamics of optimal transport \,---\,}
We now investigate the properties of the optimal protocol $\lambda_{\rm OT}(t)$, in direct comparaison with other finite-time methods.
We complement analytical results with numerical solutions of the Langevin equation [Eq.~(\ref{eq:Langevin})].
For this analysis, we adopt physical parameters and conditions typical of optical trapping experiments involving micron-sized particles at room temperature \cite{Rosales2020, Pires2023, oikawa2025experimentally}. Specifically, we set the characteristic correlation time of the particle’s position in the trap to $\omega_0^{-1} \approx 10^{-3}$\thinspace{s}, and the corresponding positional standard deviation to $\sigma = \sqrt{D / \omega_0} \approx 15$\thinspace{nm}. We study transformations where the potential is shifted from $\lambda_i = 0$ to $\lambda_f = 5 \sigma$.

In Fig.~\ref{fig:ProtocolsAndEnergetics}~(a), we present the OT protocol $\lambda_{\rm OT}(t)$ (red line) for a protocol duration set to $t_f = \omega_0^{-1}/2$. Here, the two discontinuities in the protocol and its strong overshoot beyond the target position are clearly visible \footnote{Its shape resembles previously reported optimal protocols for stiffness changing transitions \cite{Rosales2020}, but features a simpler linear evolution during the overshoot phase}.
Due to the Gaussian and linear nature of the process, the initially Gaussian distribution remains Gaussian at all times.
Furthermore, if the initial state is at thermal equilibrium, the variance remains constant in time.
The evolution of the state is therefore fully characterized by its mean $u(t) = \langle x(t) \rangle$.
In addition to the exact theory, we numerically evaluate trajectories from $10^4$ independent Langevin simulations subjected to the optimal protocol $\lambda_{\rm OT}(t)$, which is also shown in Fig.~\ref{fig:ProtocolsAndEnergetics}~(a) (gray-blue dash-dotted line). As expected, the mean position successfully relaxes to the new equilibrium value $\lambda_f$ within the prescribed time $t_f$: the protocol constrains the equilibration time of the system, below the thermal relaxation time.

The acceleration of relaxation necessarily comes at an energetic cost: work is exchanged with the time-dependent confining potential while heat is exchanged with the surrounding heat bath \cite{Sekimoto1998}.
Each protocol $\lambda(t)$ together with associated system's response $u(t)$ lead to a distinct ensemble-averaged work $W(t)$ and heat $Q(t)$ (see Appendix~\ref{App:Energetics} for details).
The First Law of thermodynamics imposes that, at the end of the process, the cumulative heat and work are equal \cite{Sekimoto1998}.
The net cost of each protocol can therefore be evaluated as the total accumulated work cost $\Delta W = \int_0^{t_f} \dot W(t) dt$.
For the OT protocol, the cost $\Delta W_{\rm OT}$ splits into three additive contributions: one from the linear part of the protocol and two from the initial and final discontinuities (Appendix~\ref{App:Energetics}). Their sum yields a remarkably compact expression,
\begin{equation}
    \Delta W_{\rm OT} = \frac{\gamma \Delta \lambda^2}{t_f}
    \label{eq:MeanOptWork}
\end{equation}
shown as a red bar plot in Fig.~\ref{fig:ProtocolsAndEnergetics}~(e).
It is the product of the mean drag force $\gamma \dot u$ and the distance $\Delta \lambda$.

The advantages of this OT protocol become particularly evident when compared to other types of driving protocols.
Without thermodynamic optimization, an accelerated relaxation can be obtained by deriving an SST protocols $\lambda_{\rm SST}(t)$ \cite{guery-odelin_driving_2023, martinez_engineered_2016, Raynal2023} for the case of a moving potential with constant stiffness.
To do so, we impose  a polynomial Ansatz for the mean position $u(t)$ and solve the EOM for $\lambda(t)$ with appropriate boundary condition (Appendix \ref{App:SST}), ensuring continuity and equilibration of $u(t)$ within the imposed time $t_f$ as seen Fig.~\ref{fig:ProtocolsAndEnergetics}~(b) \footnote{Due to the simplicity of the geodesic $u_{\rm OT}(t)$, the optimal protocol $\lambda_{\rm OT}(t)$ can also be re-derived in the spirit of an SST approach by postulating a linear evolution for the mean position. Postulating a linear evolution suffices to obtain inadvertently the optimal and constrained protocol.}.
This cost of this unoptimized SST protocol $\Delta W_{\rm SST}$ (orange bar in Fig.~\ref{fig:ProtocolsAndEnergetics}(e)) is exactly $20\%$ larger than the cost of the OT protocol.
Importantly, both OT and SST protocols yields energetic costs which are independent of the stiffness of the employed potential, which is not the case generally (Appendix~\ref{App:Optimal}).

In contrast, unconstrained OC protocols do not impose final equilibrium state but seek for a global energetic minimum \cite{Schmiedl2007, olsen2025harnessing,loos2024universal}.
Such protocol is shown in Fig.~\ref{fig:ProtocolsAndEnergetics}~(c) together with the associated mean position, which does not reach the final equilibrium mean position within the finite duration $t_f$. Instead, $u(t)$ still follows an exponential relaxation toward the final equilibrium state after the end of the driving.
Interestingly, the expression of the OC protocol, $\lambda_{\rm OC}(t) = \lambda_i +  \frac{\Delta \lambda(\omega_0 t + 1)}{\omega_0 t_f + 2}$, and its energetic cost $\Delta W_{\rm  OC} = \frac{\kappa \Delta \lambda^2}{\omega_0 t_f + 2}$ closely resemble our result in Eq.(\ref{Eq:OptProtocol}) and Eq.~(\ref{eq:MeanOptWork}).
The energetic cost of the OC protocol (green bar in Fig.~\ref{fig:ProtocolsAndEnergetics}~(e)) is significantly smaller than the cost of the OT and SST protocols, which constrain the final state of the system.

Finally, the simplest STEP protocol $\lambda_{\rm STEP}(t)$, shown on Fig.~\ref{fig:ProtocolsAndEnergetics}~(d), is both unconstrained and unoptimized.
In that case, the mean position follows an exponential decay $u(t) = \lambda_f(1 - \rm{exp}[-\omega_0 t]])$ and the work injected reads $\Delta W_{\rm STEP} = \frac{\kappa}{2} \lambda_f^2$ (blue bar Fig.~\ref{fig:ProtocolsAndEnergetics}~(e)) exceeds the cost of the OC protocol.
In the case of both unconstrained OC and STEP protocol, the work cost depends directly on the stiffness of the potential, in contrast with the cost of OT and SST solutions.\\

\begin{figure}
	\centerline{\includegraphics[width=1\linewidth]{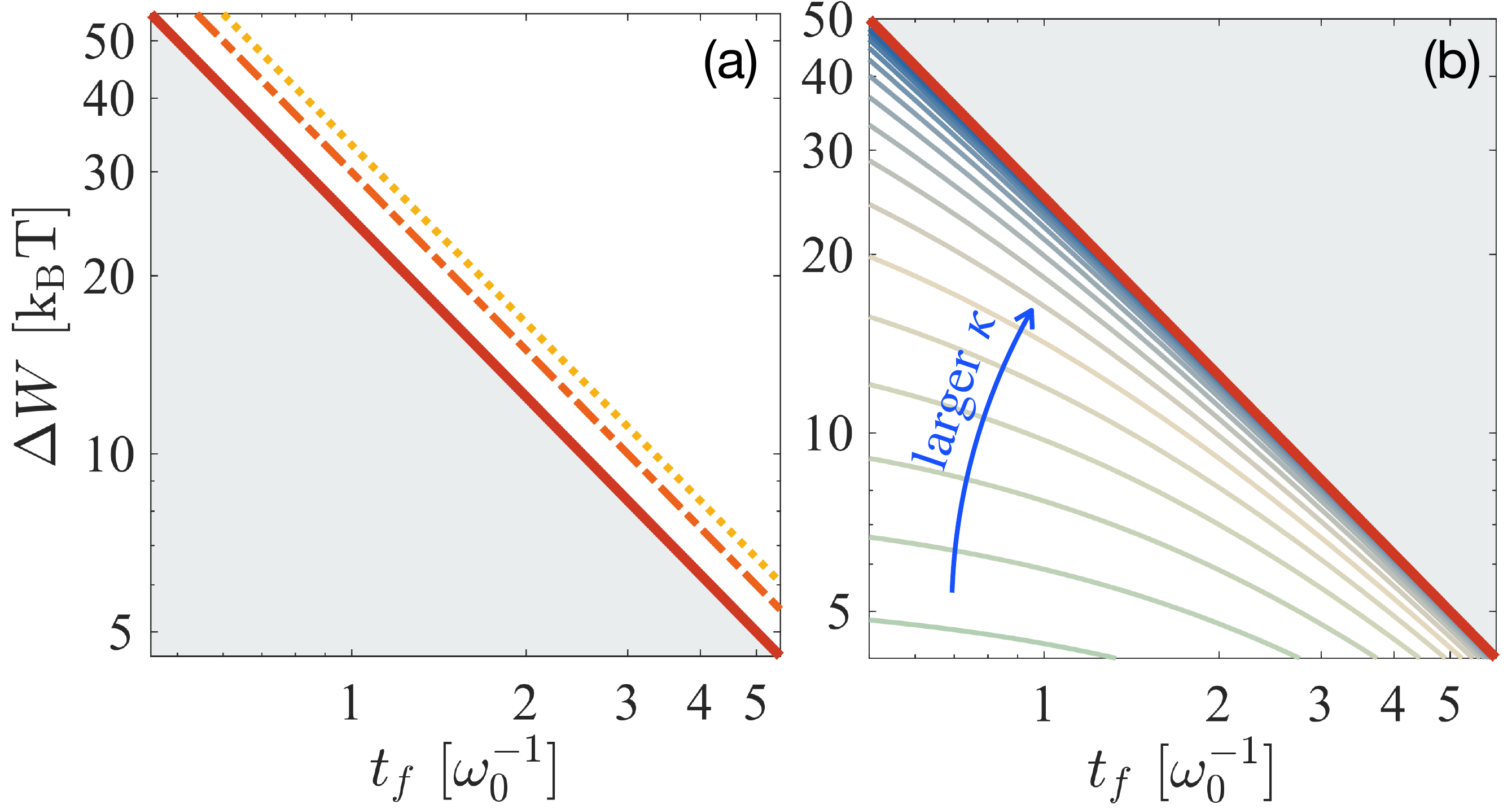}}
	\caption{
    Total work cost of the protocols in units of $k_B T$ as a function of their duration $t_f$ in units of $\omega_0^{-1}$
    (a) The cost of the optimal protocol (red solid line) is a lower bound for SST in time, excluding the lower triangular region highlighted in light gray. The cost of a third-order (orange dash-dotted line) and of a second order (golden dotted line) SST protocols are shown, both belonging to the upper triangle. (b) The same cost is an upper bound for unconstrained OC protocols (green to blue solid lines for increasing stiffness $\kappa_{\rm OC} \in [0.1 \kappa, 10\kappa]$).
    }
	\label{fig:CostVsDuration}
\end{figure}

\textit{Lower bound, upper bound}---
The OT protocol plays a key role in the thermodynamics of finite-time processes since, as seen in Fig.~\ref{fig:CostVsDuration}~(a,b) the time-energy relation $\Delta W_{\rm OT}(t_f)$ Eq.~(\ref{eq:MeanOptWork}) which is obeyed when following the OT protocol corresponds to two distinct bounds.

First, due to the optimization procedure, the cost $\Delta W_{\rm OT}$ is a lower bound for any SST protocols, which we illustrate in Fig.~\ref{fig:CostVsDuration}~(a) by comparing $\Delta W_{\rm OT}$ to $\Delta W_{\rm SSR}$ for third (orange line) and second-order (yellow line) polynomial SST solutions (Appendix \ref{App:SST}).
As a consequence of optimization, the cost of OT forms the boundary of a region (shown in gray) inaccessible to any other protocol that ensures final equilibrium within $t_f$.
Second, the same $\Delta W_{\rm OT}$ is an upper bound for OC protocols as we illustrate in Fig.~\ref{fig:CostVsDuration}~(c).
The cost $\Delta W_{\rm OC}$ depends on the stiffness, which leads to a degeneracy of time-energy relations $\Delta W(\kappa, t_f)$.
As the stiffness increases, the work $\Delta W_{\rm OC}$ converges to the work engaged in the constrained optimal protocol $\Delta W_{\rm OT}$.
For slow protocols, both work cost also meets, converging to $0$ in the quasistatic limit.
Importantly, for OC protocols, $t_f$ corresponds to the duration of the protocol, but not to the full relaxation of the system.
In the next section, we show that this double-bounding role becomes a fundamental time-energy minimum in a case where the duration of the relaxation is penalizing the process.\\

\textit{Application: optimal time-energy tradeoff for stochastic resetting \,---\,}
We now explore the application of this optimal dragging protocol to finite-time implementation of SR \cite{Evans2011, Evans2020, evans2018effects}.
A crucial result of SR lies in its ability to accelerate search processes, by reducing the MFPT needed for a particle to reach a target located at an arbitrary distance $x_0$ \cite{Evans2011}.
More precisely, for Poissonian resetting to the origin at a rate $r$, and for a target located at a position $x_0$, the MFPT $\langle \mathcal{T}_{\rm inst} \rangle$ of an instantaneous resetting process reads $\langle \mathcal{T}_{\rm inst} \rangle = r^{-1} (\exp[x_0 / \langle|x|\rangle] - 1)$ where $\langle|x|\rangle = \sqrt{D/r}$ is the mean absolute position in the non-equilibrium steady-state \cite{Evans2011}.
From a thermodynamic point of view, resetting drives the system into a non-equilibrium steady-state, maintained only via a constant supply of energy \cite{Fuchs2016, Busiello2020, Gupta2020_thermo, TalFriedman2020, Mori2022, goerlich2023experimental, olsen2024thermodynamic, gupta2025thermodynamic, abdoli2024shear, olsen2024thermodynamic2}.
In experiments, SR is implemented via explicit finite-time return protocols \cite{TalFriedman2020, Besga2020, Besga2021, goerlich2023experimental}.
In that case, the MFPT is increased by a number $n= r \langle \mathcal{T}_{\rm inst} \rangle $ of resetting events of duration $t_{\rm rst}$ that occur before the target is reached \cite{evans2018effects, TalFriedman2020, Besga2020}.
It leads to a longer MFPT $\langle  \mathcal{T} \rangle = \langle \mathcal{T}_{\rm inst} \rangle \left( 1 + r t_{\rm rst}\right)$.
Important theoretical efforts have explored the consequences of such realistic finite-time resetting \cite{evans2018effects, pal2019invariants, gupta2020stochastic, bodrova2020resetting, gupta2021resetting, gupta2022work}.
Energetically, the cost of such realistic SR processes is the sum of the cost of each return protocol \cite{TalFriedman2020, goerlich2023experimental}.
Mitigating the energetic cost of finite-time SR with its effect of the MFPT  is necessary to obtain an optimal search solution \cite{evans2013optimal, pal2023thermodynamic, singh2024emerging, sunil2023cost}.

Here, we demonstrate that the use of OT protocol to apply SR leads to an optimal time-energy bound, mitigating MFPT increase and dissipation.
To do so, we apply $\lambda_{\rm OT}$ for each resetting event, using as initial condition $\lambda_i = x(t)$ the stochastic position $x(t)$ of the particle at the resetting instant $t$ and $\lambda_f = 0$ as end-point.
This allows to drag the particle from $x(t)$ back to the origin within a finite time $t_{\rm rst}$.

The use of constrained protocols ($\lambda_{\rm OT}$ or $\lambda_{\rm SST}$ -Fig.~\ref{fig:ProtocolsAndEnergetics}~(a, b)) enables a clear control over $t_{\rm rst}$: the duration of the resetting event is imposed by the protocol $t_{\rm rst} = t_f$.
For unconstrained protocols ($\lambda_{\rm OC}$ or $\lambda_{\rm STEP}$ -Fig.~\ref{fig:ProtocolsAndEnergetics}~(c, d)), the duration of the resetting event is ill-defined.
Still, the exponential nature of the relaxation in the resetting potential allows us to evaluate the time needed to bring the average position $u(t)$ back to the origin up to a fixed threshold (that we define as $.1\%$ of the standard deviation in the resetting potential $\sigma$, more detailed is found in Appendix \ref{App:Sim} and \ref{App:FullRelax}).

From a thermodynamic point of view, the work cost of each protocol can be evaluated with stochastic thermodynamics, as shown above.
At the level of individual trajectories and following the optimal protocol derived here, the work cost of a single resetting event reads $\Delta w_{\rm OT} = \gamma x(t)^2 / t_{\rm rst}$.
If one ignores the presence of the absorbing boundary in $x = x_0$, the average work is characterized by the typical length scale $\zeta \equiv \sqrt{\langle x^2\rangle} = \sqrt{2D/r}$.
It leads to a remarkably simple result $W_{\rm OT}^{\rm rst, s} = \gamma \zeta^2/t_{\rm rst} = 2 k_B T / r t_{\rm rst}$.
Instead, taking into account the fact that the target forms an absorbing boundary excludes contributions of transports from $x>x_0$ to the origin.
This reduces the mean work to $W_{\rm OT}^{\rm rst} = 2 \alpha k_B T / r t_{\rm rst}$.
The correction factor $\alpha = 1 - (1 - e^z)^2 + \frac{z^2}{4} \left( 1 - \cosh{z \sqrt{1-e^{-z}}}\right)$, with $z = x_0 \sqrt{r/D}$, can be derived from calculating the second moment of the quasi-stationary solution to the Fokker-Planck equation with an absorbing boundary \cite{Evans2020} and is independent of $t_{\rm rst}$ \footnote{The influence of the non-zero width of the distribution after resetting, is taken into account by the measured $\mathcal{T}_{\rm inst}$ \cite{Besga2021}. The possibility that the particle reaches the target \textit{during} the resetting event is ignored. The limits of our assumptions are tested in Appendix \ref{App:Sim}.}.
This simple result yields a tradeoff relation which optimally mitigates MFPT and energetic cost
\begin{equation}
    \langle \mathcal{T}\rangle = \langle \mathcal{T}_{\rm inst} \rangle  \left( 1 + \frac{2 \alpha k_B T}{W_{\rm OT}^{\rm rst}} \right).
    \label{eq:SrBound}
\end{equation}
One can reduce the MFPT only at the price of a larger thermodynamic work, or conversely, one can obtain SR at a low thermodynamic cost, at the price of an increasing MFPT. Similar tradeoff relations have been studied in the past without explicit optimization, all of which features the generic property that a higher work leads to a shorter MFPT \cite{singh2024emerging, pal2023thermodynamic}.
The OT protocol utilized here, however, ensures the optimal tradeoff between work and MFPT.

\begin{figure}[htb]
	\centerline{\includegraphics[width=1\linewidth]{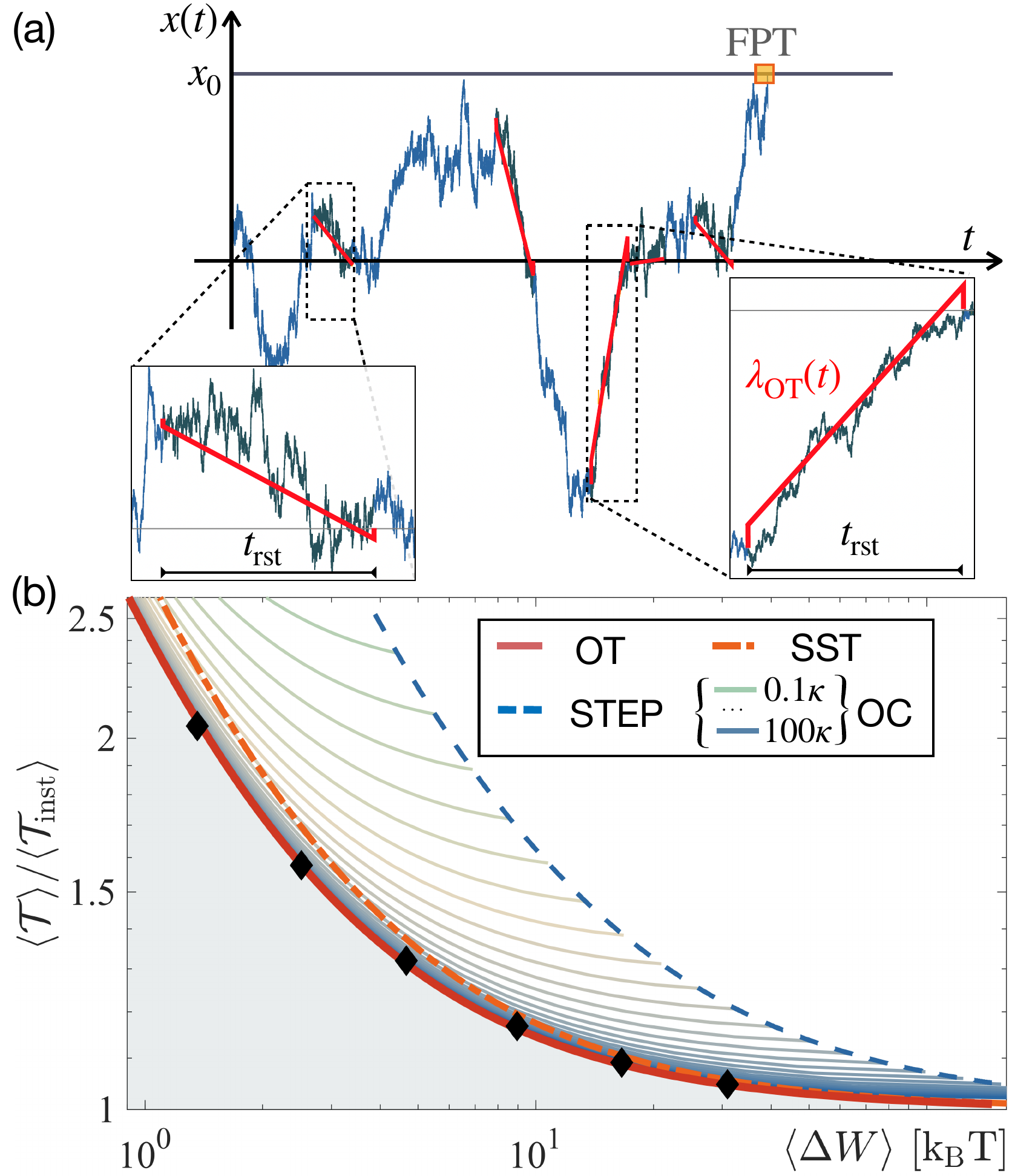}}
	\caption{(a) Typical snapshot of a numerical simulation of SR implemented with OT protocol. We magnify two examples of resetting events, with the protocol underlined in red, to highlight that both the slope and the amplitude of the jumps in the protocol are depending on the position $x(t)$ at the beginning of the resetting event. The duration of all resetting events is fixed to $t_{\rm rst}$.
    (b) Time-energy trad-off relation for resetting: mean first-passage time $\langle \mathcal T \rangle$ of finite-time resetting normalized by its instantaneous counterpart $\langle \mathcal T_{\rm inst} \rangle$ as a function of the average thermodynamic cost of each resetting event (in units of $k_B T$). MFPT and work cost of the OT protocol form a time-energy bound (the red solid line corresponds to Eq.~(\ref{eq:SrBound}) and the black diamonds to the result of numerical simulations) defining an excluded region (gray filled area). All other protocols, SST (orange dash-dotted line), STEP (blue dashed line) and unconstrained OC (green to blue solid lines for increasing stiffness) lie above this bound [Parameters: target distance $x_0 = 0.1~\rm{\mu m}$, resetting rate $r = 70 ~s^{-1}$ \textit{i.e.} the optimal rate that minimized $\langle \mathcal{T}_{\rm inst} \rangle $ \cite{evans2013optimal}, stiffness of the resetting potential $\kappa = 20 ~\rm{pN/\mu m}$ for the constrained protocols and stiffness $\kappa_{\rm OC} \in [.1, 100]\kappa$ for OC protocols].}
	\label{fig:Resetting} 
\end{figure}

The obtained tradeoff Eq.~(\ref{eq:SrBound}) is shown on Fig.~\ref{fig:Resetting}~(b) (red solid line).
It is fully characterized by $t_{\rm rst}$, independently of the potential stiffness.
The analytical result is verified with numerical simulations (black diamonds).
More generally, the work expression used in Eq.~\ref{eq:SrBound} is valid as long as the standard deviation in the resetting potential $\sigma$ is small with respect to the typical scale of the SR process $\zeta$.
This is always the case in realistic experimental contexts \cite{TalFriedman2020, Besga2020, goerlich2023experimental}.
A detailed quantitative verification of the limits of our assumptions is presented in Appendix~\ref{App:Sim}, demonstrating that our results hold even for surprisingly low stiffness, with $\sigma \approx \zeta$.

The OT protocol achieves a minimal time-energy bound for any finite-time implementation of SR, as we verify via comparison with other driving protocols.
Using SST protocol (orange dot-dashed line) yields a $20\%$ higher energetic cost for the same time-constraint.
Conversely, when using unconstrained OC protocols, the energetic cost is low, but the temporal cost is large.
In that case, the cost depends on the stiffness of the resetting potential.
This degeneracy lead to a series of time-energy lines (from green to blue thin lines) in Fig.~\ref{fig:CostVsDuration}, each corresponding to varying $t_{\rm rst}$ with a constant stiffness.
In the high stiffness (towards blue line) and/or slow protocols limits (towards small $W$), it approaches the OT protocol bound.
In the fast protocol limit, the  OC protocols reach the limit set by STEP protocols, as one expects.
We showed in Fig.~\ref{fig:CostVsDuration}~(b,c), that the optimal protocol derived here forms a twofold bound, separating constrained from unconstrained driving protocols.
In the case of SR, relaxation time becomes a variable of interest and this bound takes a stronger optimal tradeoff form, mitigating time and energy in an ideal way, as illustrated by Eq.~(\ref{eq:SrBound}) and Fig.~\ref{fig:Resetting}~(b).\\

\textit{Conclusion.}
In this work, we use a Lagrangian method to derive optimal transport protocol for the translation of a diffusive particle in a moving harmonic trap.
The protocol enforces equilibrium over finite time, which allows fast transition for the system, at a minimal energetic cost, which forms a key tradeoff in the design of micro-machines.
The cost of this optimal transport protocol forms a double-sided time-energy bound on the energetics of other families of driving protocols.
The application of this protocol to resetting further illustrates its relevance: it leads to a universal tradeoff relation between the speedup of the search process and its energetic cost, bounding from below all other protocols.
This provides a method for realistic implementations of SR mitigating efficiently energy and time.

\acknowledgements
R.G. thanks Giovanni Manfredi, Cyriaque Genet, David Gu\'ery-Odelin, Tomer Markovich, Haim Diamant and Yair Shokef for insightful discussions.
Y.R. and R.G. acknowledge support from the European Research Council (ERC) under the European Union’s Horizon 2020 research and innovation program (Grant Agreement No. 101002392).
 KSO acknowledges support from the Alexander von Humboldt foundation. HL acknowledges support by the Deutsche Forschungsgemeinschaft (DFG) within the project LO 418/29-1. 

\appendix 


\section{Detailed derivation of the optimal protocol and its energetic cost}
\label{App:Optimal}

We seek for the protocol $\lambda_{\rm OT}(t)$ that will drive the mean position from $u(0) = \lambda_i$ to $u(t_f) = \lambda_f$ in a finite time $t_f$ while minimizing the work cost (which also correspond to the dissipated heat).
The work can be expressed from the dynamics of the particle and the confining potential, using standard stochastic thermodynamics \cite{Sekimoto1998}.
The stochastic increment of work reads $dw = \partial V/\partial \lambda d\lambda = \kappa(\lambda - x)d\lambda = \kappa(\lambda - x)\dot\lambda dt$.
The net energetic cost of a transformation is evaluated as the ensemble-averaged and time-integrated stochastic work
\begin{equation}
    \Delta W = \int_0^{t_f} \kappa (\lambda - u) \dot \lambda dt.
\end{equation}
In order to derive the optimal protocol, we build the following functional
\begin{equation}
    \mathcal{J} = \int_0^{t_f} \left(  \kappa [\lambda - u] \dot \lambda + \mu [\dot u - \omega_0(\lambda - u)] \right)
\end{equation}
where the first term in the integrand correspond to the work cost, while the second correspond to the constraint that $u$, $\dot u$ and $\lambda$ are related by the equation of motion. This constraint is imposed via a Lagrange multiplier $\mu$.
The integrand therefore forms a Lagrangian $\mathcal{L}(u, \dot u, \lambda, \dot \lambda, \mu, \dot \mu)$.
The optimal protocol $\lambda$ is obtained by solving the three Euler-Lagrange equation corresponding to each variable.
The Euler-Lagrange equation arising from the $\mu$ variable is simply the equation of motion.

The Euler-Lagrange equation lead to the constraint the optimal path has to obey $\ddot u = 0$ \textit{i.e.} $u(t) = a t + b$.
Imposing here the appropriate boundary condition leads to the optimal transport path
\begin{equation}
    u_{\rm OT}(t) = \lambda_i + \frac{\lambda_f -\lambda_i}{t_f} t.
\end{equation}
Inserting this solution in the equation of motion yields the optimal transport protocol
\begin{equation}
    \lambda_{\rm OT}(t) = \lambda_i + \frac{(\lambda_f -\lambda_i)(\omega_0 t + 1)}{\omega_0 t_f}.
\end{equation}

\begin{figure}[htb]
	\centerline{\includegraphics[width=0.9\linewidth]{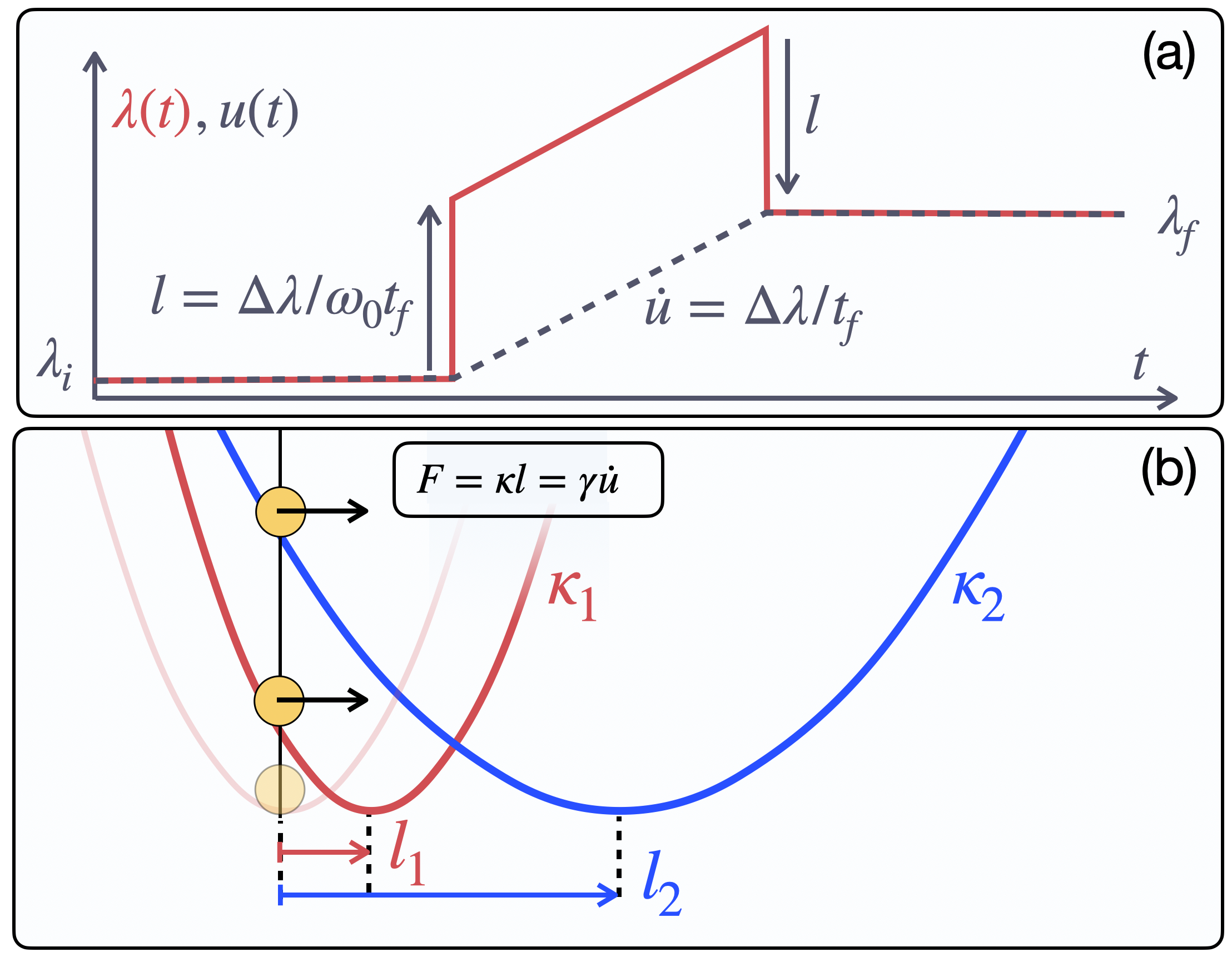}}
	\caption{(a) Schematic representation of the optimal protocol $\lambda_{\rm OT}(t)$. The jumps of amplitude $l$ in the beginning and at the end of the protocol are such as to ensure a constant force exactly canceling the viscous drag on the Brownian object. It allows the geodesic transport at a constant velocity from $\lambda_i = 0$ to $\lambda_f$ in time $t_f$.
    (b) The amplitude of the jumps $l$ are, by construction, ensuring a constant force on the Brownian particle. The jumps $l_1$ and $l_1$ are adapted to the stiffness of the potential.
    }
	\label{fig:SchemaProtocol}
\end{figure}

In Fig.~\ref{fig:SchemaProtocol}~(a) we sketch the optimal protocol (red line) as well as the mean trajectory $u(t)$ responding to it.
We underline with vertical arrows the two discontinuities, of amplitude $l = \Delta\lambda/\omega_0 t_f$.
During the protocol, the particle moves at constant speed $\dot u = \Delta\lambda/t_f$.
In Fig.~\ref{fig:SchemaProtocol}~(b) we illustrate the dependence of the amplitude of the jumps $l$ on the stiffness.
The discontinuities impose a constant force $\kappa l$ immediately balancing the drag force $\gamma \dot u$.
This is the origin of the stiffness independence of the work cost, in variance with unconstrained protocols.

\section{Relation to geometric approaches}
\label{App:Geom}

The time-dependent probability density $P(x,t)$ of the Brownian particle in the potential $V(x,t)$ obeys the Fokker-Planck equation $ \partial_t P(x,t) = - \partial_x [v(x,t)P(x,t)] $ where $v(x,t) = - \partial_x[V(x,t) + k_B T \ln P(x,t)]/\gamma$ is the mean local velocity.
Optimal transport seeks to find the time-dependent potential $V_{\rm OT}(x,t)$ which will enforce the optimal velocity field and probability density $\left\{ v_{\rm OT}(x,t), P_{\rm OT}(x,t) \right\}$ such that the entropy production $\Sigma$ is minimized, under the constraint that the system obeys the Fokker-Planck equation \cite{villani2008optimal, dechant2019thermodynamic}.
The total entropy production $\Sigma = \Delta S_{\rm sys} + \Delta Q/T$ is the sum of the system's entropy $\Delta S_{\rm sys}$ and the medium's entropy given by the heat exchanges with the bath divided by the temperature $\Delta Q/T$ \cite{Seifert2005, seifert2012}.
The Second Law sets its positivity $\Sigma \geq 0$.
It can be expressed as the integral $\Sigma(t) = \int_0^{t} \sigma(t') d't$ of the entropy production rate, which itself can be written using $v(x,t)$ and $P(x,t)$ \cite{SekimotoBook, seifert2012}
\begin{equation}
    \sigma(t) = \frac{\gamma}{T} \int |v(x,t)|^2 P(x,t) dx.
\end{equation}

The minimal entropy production $\Sigma_{\rm OT}$ for a transformation from $P_i(x)$ to $P_f(x)$ within a finite time $t_f$ is reached if the system's evolution $P(x,t)$ follows, at constant speed, a geodesic of the space spanned by the $L^2$-Wasserstein distance \cite{aurell2012refined, nakazato2021geometrical}.
Its value is then given by
\begin{equation}
    \Sigma_{\rm OT} = \frac{\gamma}{T t_f} \mathcal{W}^2(P_i, P_f).
\end{equation}
In our case of a moving Gaussian in a harmonic oscillator of constant stiffness and as demonstrated in Refs. \cite{dechant2019thermodynamic, nakazato2021geometrical}, $\mathcal{W}^2(P_i, P_f) = |\lambda_f - \lambda_i|^2$ and we obtain $\Sigma_{\rm OT} = \frac{\gamma \Delta \lambda^2}{T t_f}$.
Since in that case the system's entropy is constant $\Delta S_{\rm sys}=0$, we have $\Delta Q_{\rm OT} = \Delta W_{\rm OT} = T \Sigma_{\rm OT} = \gamma \Delta \lambda^2/t_f$ which corresponds to the cost of the OT protocol given in the main text.

The time-dependent driving $U_{\rm OT}(t) = \kappa [x - \lambda_{\rm OT}(t)]^2/2$ is determined by the protocol $\lambda_{\rm OT}(t)$ which decomposes into a geodesic $u_{\rm OT}(t) = \lambda_i +\Delta \lambda~ t / t_f$ and a counterdiabatic term $l = \Delta\lambda / \omega_0 t_f$ \cite{zhong2024beyond}.
The linear nature of the optimal path $u_{\rm OT}(t)$ is a consequence of the Euclidean space induced by the Wasserstein metric on Gaussian states: geodesics are straight lines \cite{dechant2019thermodynamic, nakazato2021geometrical}.
The counterdiabatic term induces two discontinuities, at the beginning and at the end of the protocol \cite{faure2025active}.
The amplitude $l$ of the discontinuities, set by $t_f$ and the physical parameters of the system, induce a force $\kappa l = \gamma \dot u$, exactly compensating the drift force.
This allows the system to follow the linear geodesic at constant speed.

\section{Energetics}
\label{App:Energetics}

\begin{figure} [h]
	\centerline{\includegraphics[width=1\linewidth]{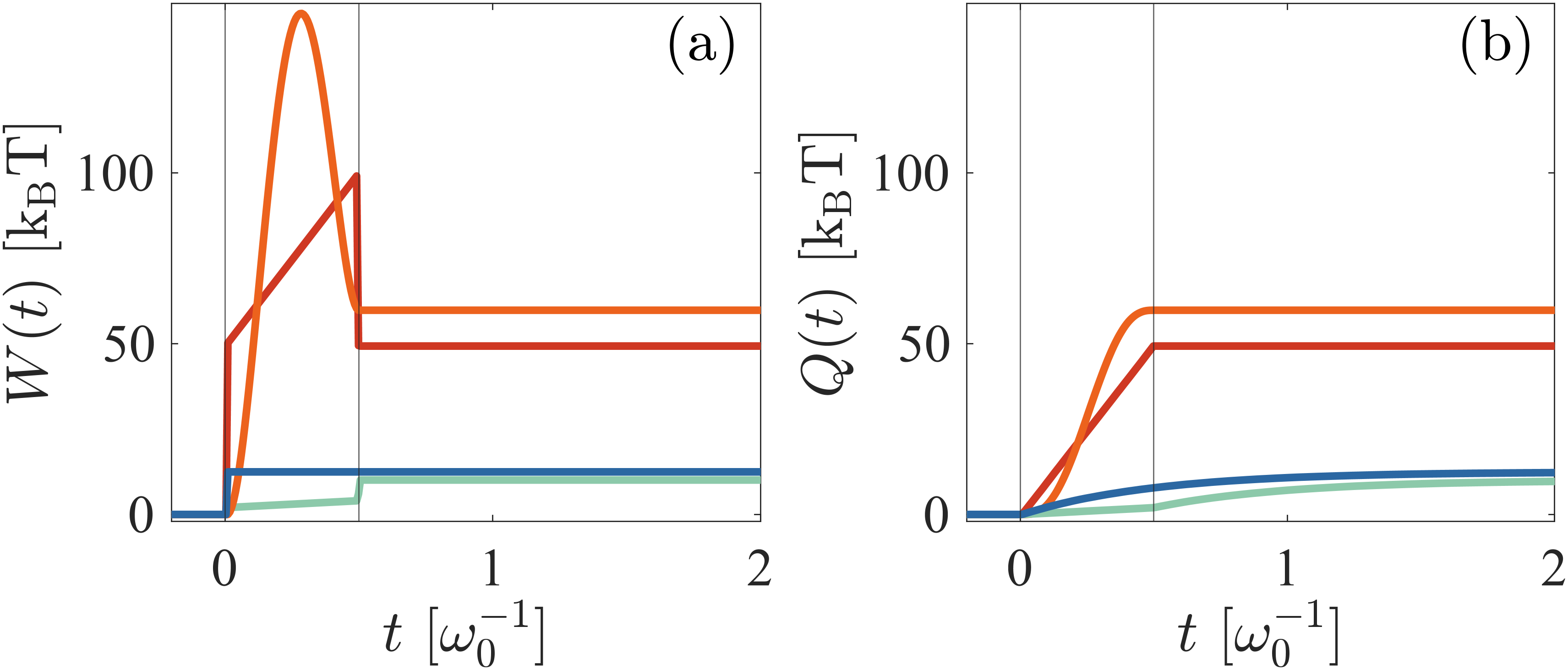}}
	\caption{Energetic cost corresponding to each protocols (same parameters as in Fig.~\ref{fig:ProtocolsAndEnergetics}). 
    (a) Mean work for protocol corresponding to OT (red), SST (orange), OC (light green) and STEP (Blue)
    (b) Mean heat in the same cases.
    }
	\label{fig:Energetics}
\end{figure}

Both work $W(t) = {\kappa} \int_0^t \dot \lambda  (\lambda - u) dt'$ and heat $Q(t) = \kappa \int_0^t \dot u  (\lambda - u) dt'$ can be measured along stochastic trajectories and averaged over the ensemble.
The time-dependent energetics under the different protocols presented in the main text is shown in Fig.~\ref{fig:Energetics}~(a,b).
The work takes a form close to the protocol's profile, while the heat follows the response of the system.
On can note in Fig.~\ref{fig:Energetics}~(b) the linear heat dissipation during the OT protocol, reflecting its geodesic nature.

The First Law imposes that $\Delta W = W(t\gg\omega_0^{-1}) - W(0) = \Delta Q$.
The net energetic cost of each protocol can be evaluated as the finite difference $\Delta W$.

\begin{figure}[htb]
	\centerline{\includegraphics[width=0.95\linewidth]{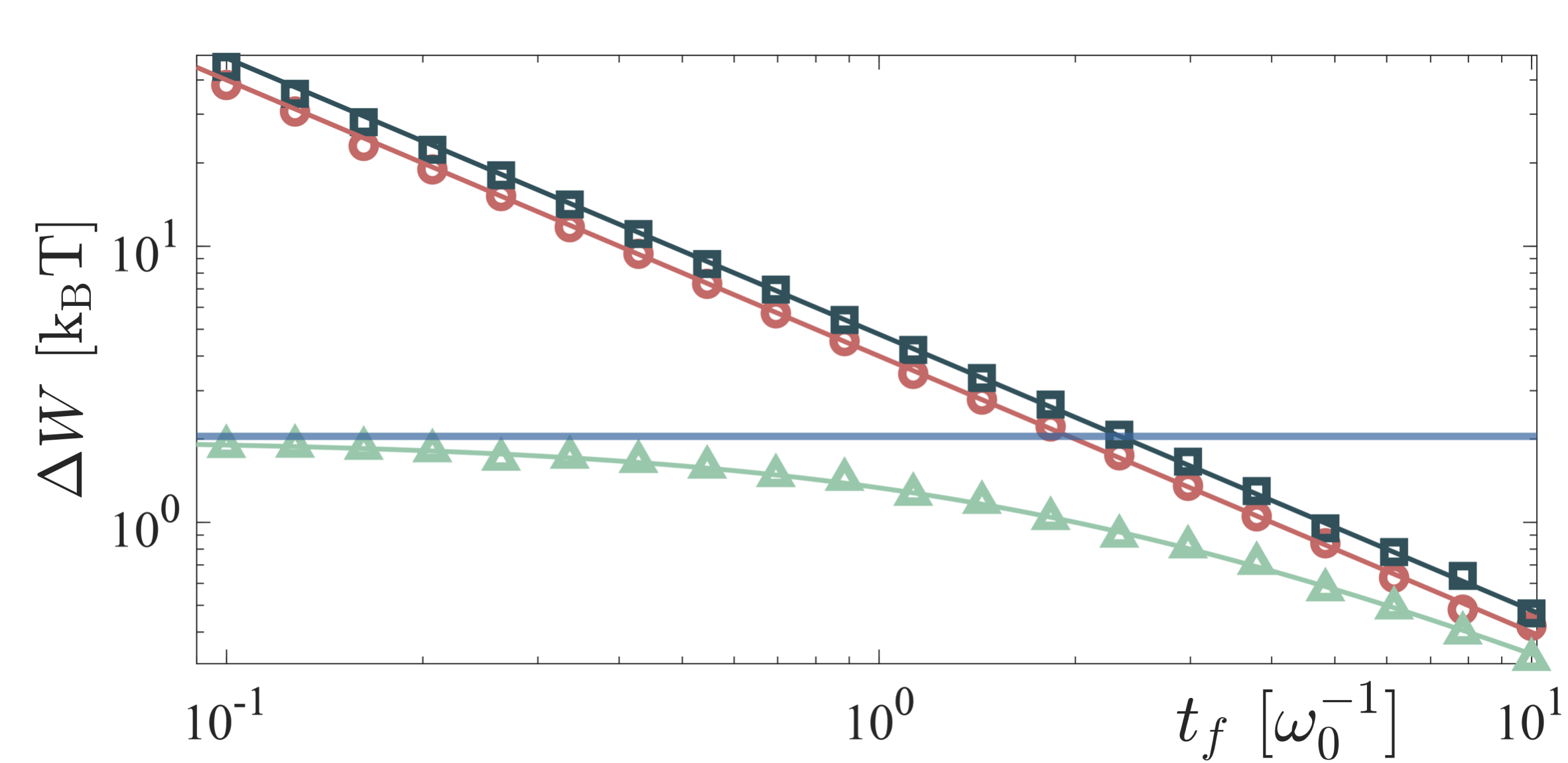}}
	\caption{Mean work as a function of the protocol duration for constrained OT (red), constrained SST (gray-blue), unconstrained OC (light-green) and STEP (blue). The symbols correspond to an average over $10^4$ Langevin simulations and the solid lines show the respective analytical predictions.
    }
	\label{fig:ProtocolSim}
\end{figure}

For the OT protocol, $\Delta W_{\rm OT}$ splits into three contributions.
The initial discontinuity from $\lambda(t=0^-) = 0$ to $\lambda(t=0^+) = l$ yields a work $\Delta W_1 = \kappa l^2 /2 $.
The protocol itself where $\lambda_{\rm OT}$ and $u_{\rm OT}$ evolve linearly yields a work $\Delta W_2 = \kappa l^2 \omega_0 t_f$.
Finally, the final discontinuity from $\lambda(t=t_f^-) = 2l$ to $\lambda(t=0^+) = \lambda_f$ yields a work $\Delta W_3 = -\frac{\kappa}{2}l^2$ that cancels the cost of the first discontinuity.

The net cost of the optimal protocol therefore reads $\Delta W_{\rm OT} = \gamma (\lambda_f - \lambda_i)^2 / t_f$ (Eq.~\ref{eq:MeanOptWork} in the main text).
As sketched Fig.~\ref{fig:SchemaProtocol}~(b) the amplitude $l$ of the discontinuities ensures a constant force $F = \kappa l = \gamma \lambda_f/ t_f$ and the optimal work takes the very simple form of $F\cdot\lambda_f$, independently of the stiffness.

In Fig.~\ref{fig:ProtocolSim}, we show the value of $\Delta W$ measured as an ensemble average over $10^4$ independent numerical simulation for various protocol durations $t_f$, with a fixed stiffness $\kappa = 1~\rm{[pN/\mu m]}$.
The results of simulations (symbols) are compared to the analytical expressions $\Delta W_{\rm OT} = \gamma \lambda_f^2/t_f$ (red), $\Delta W_{\rm SST} = 1.2 \times \gamma \lambda_f^2/t_f$ (gray-blue), $\Delta W_{\rm OC} = \kappa \lambda_f^2/(\kappa t_f/\gamma + 2)$ (light green) and $\Delta W_{\rm STEP} = \kappa \lambda_f^2/2$ (blue).

\section{Variations of engineered equilibration}
\label{App:SST}

We derive here swift state-to-state transformation protocols \cite{guery-odelin_shortcuts_2019, guery-odelin_driving_2023, martinez_engineered_2016, Chupeau2018, Raynal2023} adapted to our case of a moving harmonic potential of constant stiffness.
To do so, we impose a polynomial form on the evolution of the mean position  $u(t) = a \bar t^3 + b \bar t^2 + c \bar t + d$ where the time-variable $\bar t = t/t_f$ is normalized by the imposed protocol duration \cite{martinez_engineered_2016}.
The coefficients of the polynomial are obtained by imposing four conditions, $u(0) = 0$, $u(\bar t = 1) = \lambda_f$, $\dot u(0) = 0$ and $\dot u(\bar t = 1) = 0$, leading to the following accelerated path
\begin{equation}
    u(\bar t) =  \lambda_f \left( -2 \bar t^3 + 3 \bar t^2 \right)
    \label{eq:EsePath1}
\end{equation}
which can be readily inserted in the equation of motion Eq.~\ref{eq:EOM} to obtain the associated protocol
\begin{equation}
    \lambda_{\rm SST}(t) =  \lambda_f \left( -2 \bar t^3 + \left[3 -\frac{6}{\omega_0t_f}\right]\bar t^2 + \frac{6}{\omega_0t_f} \bar t\right).
    \label{eq:EseProtocol1}
\end{equation}

\begin{figure}
	\centerline{\includegraphics[width=0.95\linewidth]{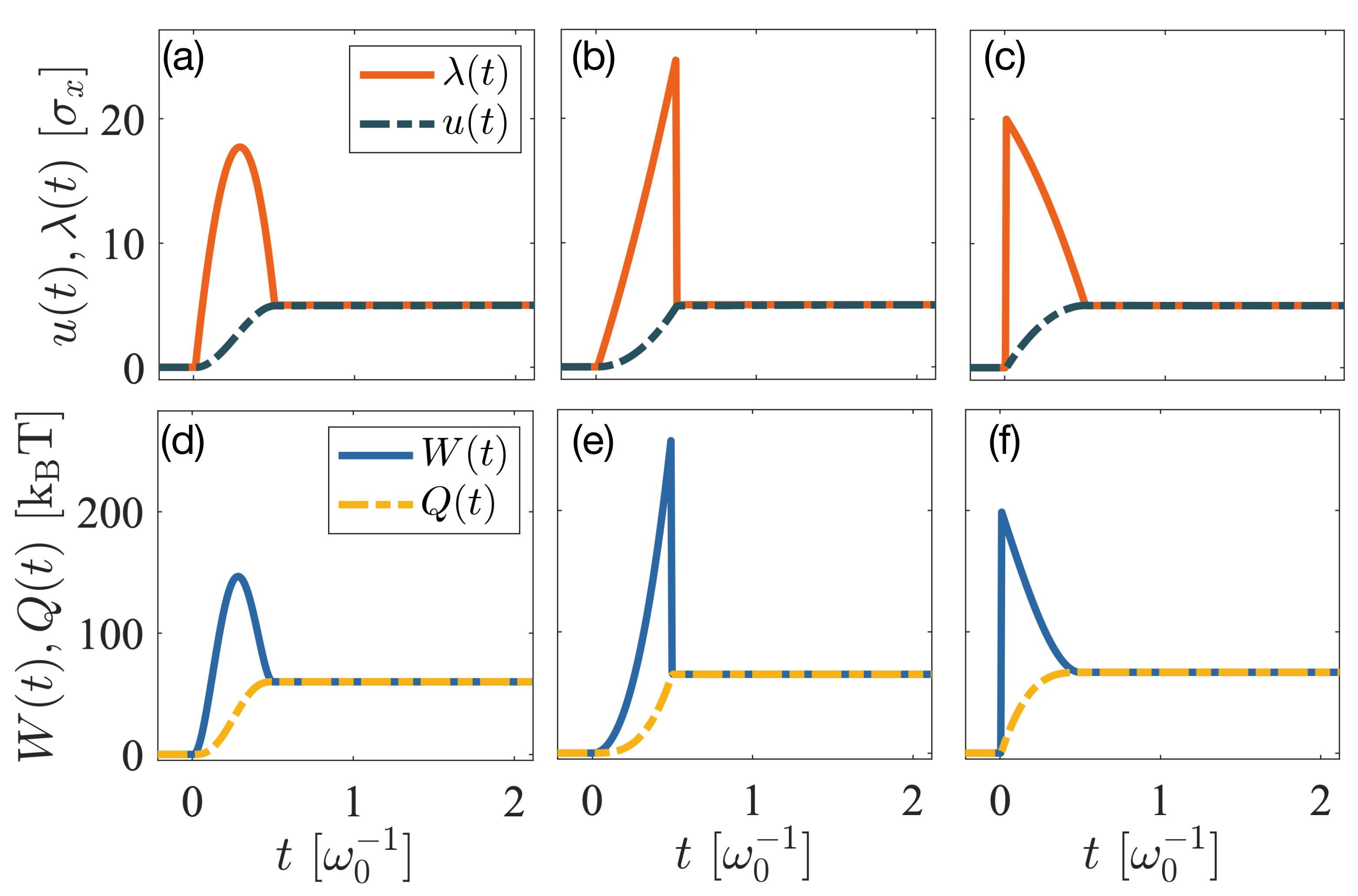}}
	\caption{(a, b, c) protocol $\lambda(t)$ (orange solid line) and mean position $u(t)$ (gray-blue dash-dotted line) in units of the standard deviation $\sigma_x$, following the three types of swift equilibration protocols Eqs.~(\ref{eq:EseProtocol1}, \ref{eq:EseProtocol2}, \ref{eq:EseProtocol3}). The mean position $u(t)$ is the average over $10^4$ Langevin trajectories undergoing the same protocol.
    (d, e, f) associated work (blue line) and heat (golden dash-dotted line) in units of thermal energy $k_B T$.}
	\label{fig:AllEse}
\end{figure}

This protocol, together with the associated path Eq.~(\ref{eq:EsePath1}) measured on an ensemble of $10^4$ Langevin simulation, is shown Fig.~\ref{fig:AllEse}~(a) (as well as Fig.~\ref{fig:ProtocolsAndEnergetics}~(b) of the main text). As clearly seen in these figures, the protocol indeed brings the mean position of the particle to the final position $\lambda_f$ within the imposed duration $t_f$.

This swift equilibration protocol is not built with a thermodynamic optimization procedures. In turn, it involves a larger amount of work and heat, in order to obtain the same acceleration as the optimized protocol.
The work injected during the protocol (shown on Fig.~\ref{fig:AllEse}~(d))  can be evaluated analytically as $W(t) = \kappa \int_0^{t_f} \dot \lambda(t) (\lambda(t) - u(t)) dt $. When integrated over the full protocol, it yields to $ \Delta W_{\rm SST} = \frac{6}{5} \Delta W_{\rm OT}$.

Due to the overdamped nature of the process, the two constraints on the final velocity $\dot u(0) = 0$ and $\dot u(1) = 0$ can be lifted.
Therefore, a lower-order polynomial can be used to build SST protocols (shown Fig.~\ref{fig:AllEse}~(b,c)).
Using a polynomial of order 2 $u(\bar t) = a\bar t^2 + b\bar t + c$ two distinct SST protocols can be derived, using three out of the four aforementioned constraints.

First, using $u(0) = 0$, $u(\bar t = 1) = \lambda_f$ and $\dot u(0) = 0$, we obtain
\begin{equation}
    \lambda_{\rm SST,2}^{(a)}(\bar t) = \lambda_f \left( \bar t^2 + \frac{2}{\omega_0 t_f}\bar t \right).
    \label{eq:EseProtocol2}
\end{equation}
Second, using $u(0) = 0$, $u(\bar t = 1) = \lambda_f$ and $\dot u(\bar t = 1) = 0$, we obtain
\begin{equation}
    \lambda_{\rm SST,2}^{(b)}(\bar t) = \lambda_f \left( -\bar t^2 + 2 \left[1 - \frac{1}{\omega_0 t_f}\right] \bar t + \frac{2}{\omega_0 t_f} \right).
    \label{eq:EseProtocol3}
\end{equation}
These protocols is shown together with their respective mean position $u(t)$ in Fig.~\ref{fig:AllEse}~(b, c).
Both second-order swift equilibration protocols present discontinuities located at the beginning or at the end of the protocol, a consequence of the choice of lifted constraint.

The energetics of these two protocols are shown in Fig.~\ref{fig:AllEse}~(e, f).
For both protocols, the total integrated work reads $ \Delta W_{\rm SST, 2}^{(a)} = \Delta W_{\rm SST, 2}^{(b)} = \frac{4}{3} \Delta W_{\rm OT}$, larger than both the optimal protocol and the third-order swift equilibration protocol.
Interestingly, the simplest case of first-order polynomial protocol retrieves the linear optimal protocol.

Constrained unoptimized driving can be realized with various protocols beyond the polynomial forms explored here.
Namely, based on such a polynomial Ansatz, the mean path $u(t)$ can be multiplied by any time-dependent function starting in $0$ and ending in $\lambda_f$.
This could serve as a basis of numerical optimization using evolutionary algorithm or machine learning by adding perturbations on a swift equilibration protocols and select cases where energetic cost is minimized \cite{casert2024learning, rengifo2025machinelearningapproachfast}.

\section{Numerical simulations of stochastic resetting with OT protocol}
\label{App:Sim}

\begin{figure}[htb!]
	\centerline{\includegraphics[width=1\linewidth]{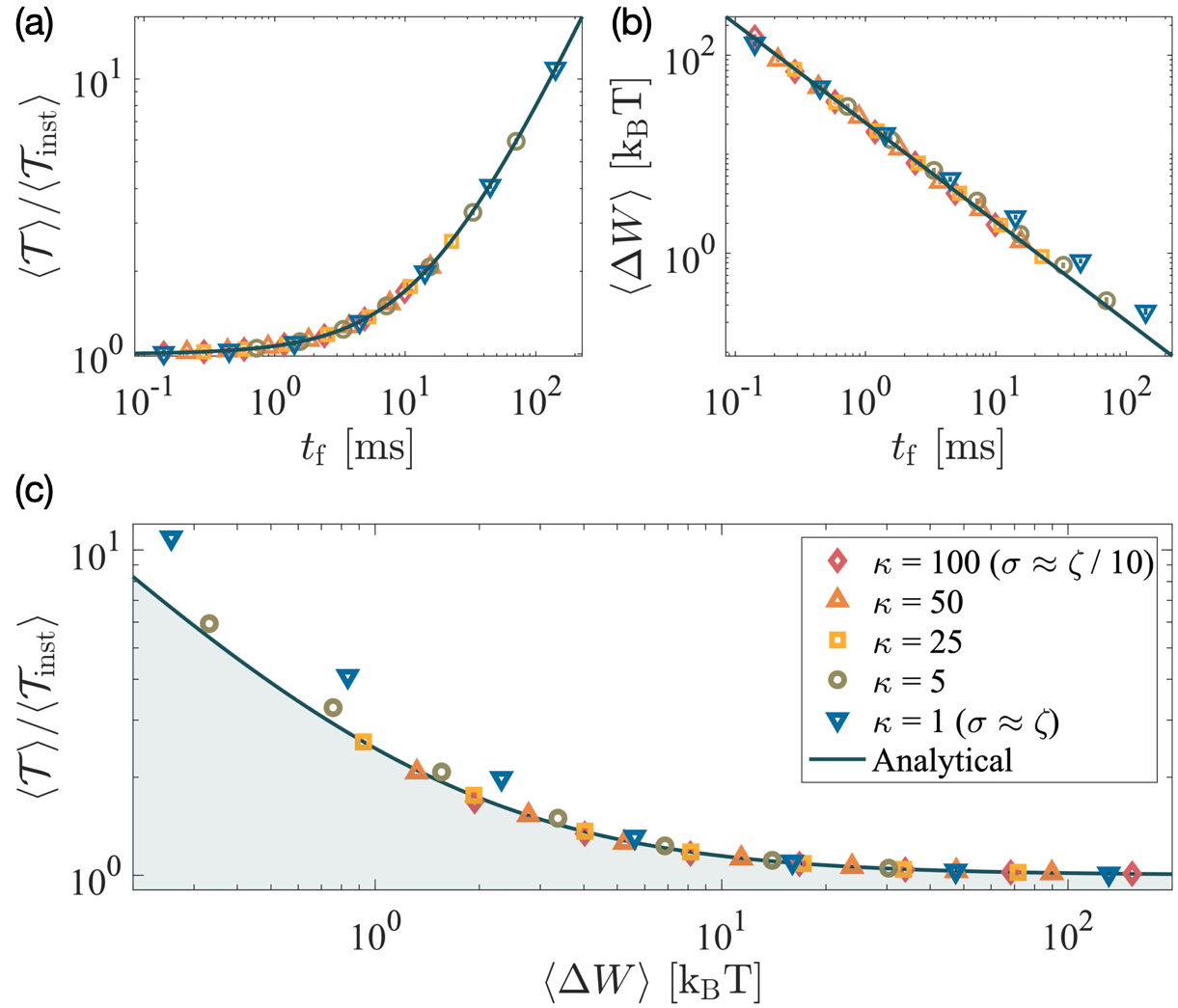}}
	\caption{Agreement between analytical results and numerical simulations, for various stiffnesses $\kappa\in[1, 100] ~\rm{pN/\mu m}$ (which is the typical range of experimentally achievable stiffnesses \cite{goerlich2023experimental}). This corresponds to cases ranging from  $\sigma \approx \zeta/100$ (red diamonds) to $\sigma\approx\zeta$ (blue down triangles).
    (a) Normalized MFPT $\langle \mathcal T \rangle / \langle \mathcal T_{\rm inst} \rangle$ as a function of the protocol duration $t_{\rm f}$, measured on numerical simulations over $N = 1500$ independent stochastic trajectories, together with the analytical result $(1 + r t_{\rm f})$.
    (b) Mean work cost of a single resetting event in units of $k_B T$ as a function of the protocol duration $t_{\rm f}$, measured on the same simulations, together with the analytical result $\langle \Delta W \rangle = 2 \alpha k_B T / r t_{\rm f}$.
    Deviations are visible for small stiffnesses, as expected.
    (c) MFPT-Work tradeoff relation Eq.~(\ref{eq:SrBound}).}
	\label{fig:SimDetail}
\end{figure}

We compare quantitatively the mean work expression $\langle W_{\rm OT}^{\rm rst} \rangle = 2 \alpha k_B T / r t_{\rm rst}$ and the tradeoff relation Eq.~(\ref{eq:SrBound}) to Brownian simulation with stochastic resetting at a rate $r$, implementing each resetting event by an OT protocol that connects $\lambda_f = x(t)$ to $\lambda_f = 0$ in a finite time $t_f$.

We expect agreement when $\sigma$ is smaller than $\zeta$ [where $\sigma = \sqrt{k_B T/\kappa}$ is the standard deviation in the potential used to apply resetting and $\zeta = \sqrt{2D/r}$ is the SR steady-state characteristic length] but show here that reasonable agreement is obtained beyond this regime.
In Fig.~\ref{fig:SimDetail} we show the result of numerical simulations (symbols) and the analytical expressions (solid lines) for (a) the MFPT, (b) the mean work per resetting and (c) the tradeoff relation.

As seen in Fig.~\ref{fig:SimDetail}~(a), normalizing the measured times by $\langle \mathcal T_{\rm inst} \rangle$ allows a perfect agreement with theory.
As seen in Fig.~\ref{fig:SimDetail}~(b) and (c), the work and tradeoff relation is also reproduced with very good precision for $\kappa \geq 25~\rm{[pN/\mu m]}$ while deviations appear for smaller stiffness.
It seems noteworthy to us that the stiffness can be reduced to rather low values, proving the experimental feasibility of this method.
Even in the case where $\sigma \approx \zeta$ (blue down triangles) deviations are only observed for slow protocols.\\

\section{Full relaxation time and threshold}
\label{App:FullRelax}

\begin{figure}[htb!]
	\centerline{\includegraphics[width=0.9\linewidth]{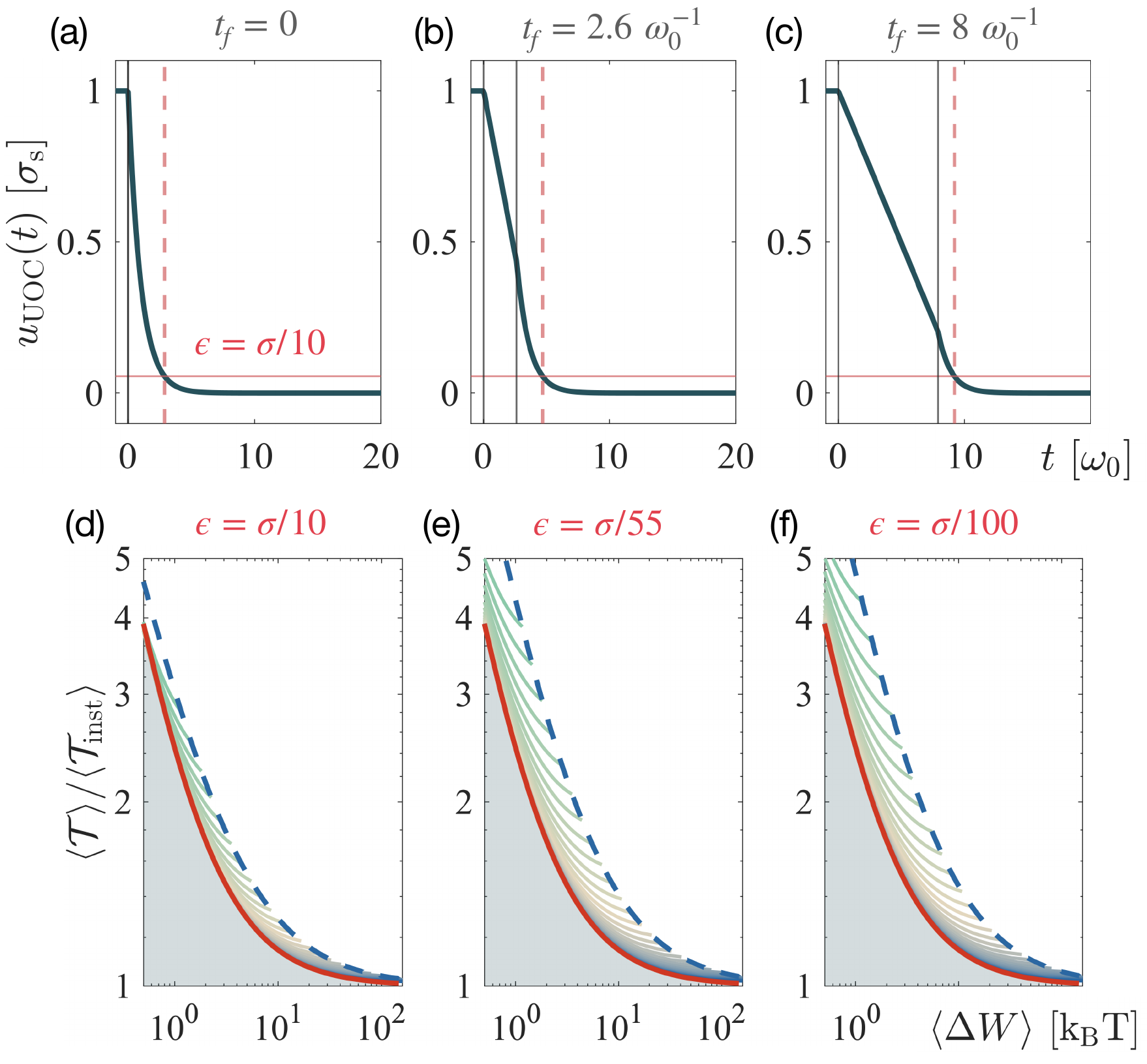}}
	\caption{Upper panels: for one choice of threshold $\epsilon = \sigma_{\rm eq}/10$, we show three cases of response to an UOC protocol, for three different protocol duration (a) $t_f = 0$, (b) $t_f = 2.6 ~\omega_0^{-1}$, (c) $t_f =8 ~\omega_0^{-1}$
    Lower panels: time-energy tradeoff for resetting as shown in the main text. The OT protocol (red line) forms a bound for UOC protocols (blue to green thin lines) and the STEP protocol (blue dashed line). We show the result for three values of the threshold, (a) $\epsilon = \sigma_{\rm eq}/10$, (b) $\epsilon = \sigma_{\rm eq}/55$, (c) $\epsilon = \sigma_{\rm eq}/100$. It shows that, below a reasonable value of the threshold, the result do not significantly evolve anymore (from b to c).
    }
	\label{fig:Threshold1}
\end{figure}

Using the unconstrained protocols, such as the UOC to implement resetting demands a definition of the time needed to achieve resetting, \textit{i.e.} to bring the position back to the origin.
Since this is an exponentially relaxing process, this time is infinite, but well captured by finite characteristic times.
We show here in detail the influence of the definition of this time on the bound proposed for resetting performances in the main text.
In Fig.~\ref{fig:Threshold1} (upper panels), we show how the resetting duration is defined: the time needed to bring the average position $u(t)$ close to the origin, up to a threshold which is defined as a by of the equilibrium standard deviation in the resetting potential $\sigma_{\rm eq} = \sqrt{k_BT/\kappa}$.
We display three cases of protocol duration, for on value of the threshold $\epsilon = \sigma_{\rm eq}/10$, which is, as is visible on the graph a quite crude approximation. It is reasonable to set threshold not larger than such example.
In Fig.~\ref{fig:Threshold1} (lower panels), we show the resetting time-energy bound for various values of the threshold.
It shows that, even up to large threshold Fig.~\ref{fig:Threshold1}~(d), the bound holds, and the UOC time-energy relations are larger than the OT result.
This confirms the results shown in the main text, which uses a fine threshold $\epsilon = \sigma_{\rm eq}/1000$.

\end{document}